\documentclass[twocolumn,aps,prd,epsf]{revtex4}
\usepackage{epsfig}
\usepackage{dcolumn}

\begin{document}

\title{Mesons and tachyons with confinement and chiral restoration, and NA60}
\author{P. Bicudo}
%\email{bicudo@ist.utl.pt}
\affiliation{Dep. F\'{\i}sica and CFTP, Instituto Superior T\'ecnico,
Av. Rovisco Pais, 1049-001 Lisboa, Portugal}
\begin{abstract}
In this paper the spectrum of quark-antiquark systems, including light
mesons and tachyons, is studied in the true vacuum and in the chiral
invariant vacuum. The mass gap equation for the vacua and the Salpeter-RPA 
equation for the mesons are solved for a simple chiral invariant and confining 
quark model.
At T=0 and in the true vacuum, the scalar and pseudoscalar, 
or the vector and axial vector are not degenerate, and in the chiral limit,
the pseudoscalar groundstates are Goldstone bosons.
At T=0 the chiral invariant vacuum is an unstable vacuum, decaying through an infinite 
number of scalar and pseudoscalar tachyons.  
Nevertheless the axialvector and vector remain mesons, with real masses.
To illustrate the chiral restoration, an arbitrary path between the 
two vacua is also studied. Different families of light-light and
heavy-light mesons, sensitive to chiral restoration, are also studied.
At higher temperatures the potential must be suppressed, and the chiral 
symmetry can be restored without tachyons, but then all mesons have small 
real masses. Implications for heavy-ion collisions, in particular for the
recent vector meson spectra measured by the NA60 collaboration, are discussed.
\end{abstract}

\maketitle

\twocolumngrid

\section{Introduction}

Very recently, the precise di-muon measurement in heavy ion indium-indium collisions
by NA60 
\cite{NA60_1,NA60_2,NA60_3,NA60_4,NA60_5}
collaboration provided an exceptional probe to observe vector mesons
in excited vacua. The masses of vector mesons in excited vacua, 
have been extensively modelled, with different results, since Brown and Rho
\cite{Brown1,Brown2,Brown3,Brown4}
proposed the scaling of the light-light vector mesons with the restoration of chiral 
symmetry. Notice that tachyons may also occur.
When there are only mesons in the vacuum, the vacuum is a minimum. It is then
stable when the minimum is absolute, or metastable when the minimum is
local because then the vacuum can decay through tunnelling. However when both mesons and 
tachyons occur, the vacuum is a saddle point. The tachyons indicate the decay directions
of the vacuum, and thus the vacuum is unstable, it is  a false vacuum. 
Here I compare the mesons and tachyons in the false chiral
invariant vacuum and in the true vacuum, in the framework of a chiral invariant
and confining potential.

Notice that, in the true vacuum of QCD, quarks are confined. On the other hand,
in the excited vacuum, chiral restoration is expected. Therefore a framework 
with a confining and chiral invariant quark interaction is convenient to study
mesons and tachyons in the two vacua.
The present study, with confinement, upgrades our knowledge of vacua and of vacuum 
fluctuations in hadronic models. For instance vacua properties of the non-confining 
sigma model \cite{Schwinger:1957em,Gell-Mann:1960np,Scadron:2006mq}
and Nambu and Jona-Lasinio model
\cite{Nambu}
have been explored in detail
\cite{Hiller},
including suprising unstabilities led by the 't Hooft $U_A(1)$ breaking
determinant
\cite{'t Hooft}. 
In the simplest scenarios, the vacua manifold of these models has
the well known Mexican hat shape, where the chiral invariant unstable 
vacuum has a finite number of tachyons.
The tachyons in the flavour SU(2) sigma model occur in the scalar 
$\sigma$ and in the pseudoscalar $ \pi^+, \ \pi^0 , \ \pi^-$ channels. 
So in the sigma model there are four tachyons in the false
chiral invariant vacuum, while in the true chiral symmetry breaking
vacuum there is one massive meson, the scalar $\sigma$ and three
pseudoscalar mesons $ \pi^+, \ \pi^0 , \ \pi^-$. In the chiral limit
the pseudoscalar mesons are goldstone bosons, in the borderline
between mesons and tachyons. 

However, when quarks suffer a confining potential, the tachyon structure of 
the false chiral invariant vacuum possibly differs from the one of the sigma 
model or the one of the Nambu and Jona-Lasinio model. In the true vacuum, 
the confining quark models have an infinite number of states in each 
channel, while the sigma model of the Nambu and Jona-Lasinio model 
only have a finite number of mesons. Other differences also occur. 
Le Yaouanc et al. 
\cite{Yaouanc2}
found that, even at high temperatures, the confining 
potential prevents a phase transition from the chiral symmetry breaking 
vacuum to the chiral invariant vacuum. 
Le Yaouanc, Oliver, Ono, P\`ene  and Raynal,
\cite{Yaouanc}
also found that, 
with a harmonic confinement, there is an infinite tower of excited vacua,
interpolating between the true chiral symmetry breaking vacuum to
the highest chiral invariant vacuum. This result was recently
generalized to any confining potential by PB and Nefediev
\cite{Bicudo_rep}.
The existence of tachyons in the chiral invariant vacuum of a
confining quark model was already signalled by 
Le Yaouanc, Oliver, Ono, P\`ene  and Raynal,
\cite{Yaouanc}.
Here these tachyons are studied in detail.

Because the present problem is quite technical, and because it is not clear
yet what is the best chiral invariant and confining quark model, for clarity 
I now use the framework of the simplest confining and chiral invariant quark model
\cite{Yaouanc,Bicudo_thesis,Bicudo_scapuz}.

Notice that a calibration problem exists in chiral computations.
The full hadron spectrum remains to be correctly reproduced. 
When the quarks were discovered, it was realized 
that the main difficulty of the quark model consisted in understanding the low 
pion mass. But Nambu and Jona-Lasinio
\cite{Nambu}
had already shown that the spontaneous dynamical breaking of
global chiral symmetry provides a mechanism for the generation
of the constituent fermion mass and for the almost vanishing
mass of the pion. This mechanism was extended to the quark model 
by le Yaouanc, Oliver, Ono, P\`ene and Raynal with the Salpeter equations
in Dirac structure 
\cite{Yaouanc}
and by PB and Ribeiro with the equivalent Salpeter equations in a form
\cite{Bicudo_thesis}
identical to the Random Phase Approximation (RPA) equations
of Llanes-Estrada and Cotanch
\cite{Llanes-Estrada_thesis}.
These chiral quark models also comply with the PCAC theorems, 
say the Gell-Mann Oakes and Renner relation
\cite{Yaouanc,Bicudo_scapuz},
the Adler Zero
\cite{Bicudo_PCAC,weall_pipi,Bicudo_piN},
the Goldberger-Treiman Relation 
\cite{Delbourgo,Bicudo_PCAC},
or the Weinberg Theorem
\cite{weall_pipi,Bicudo_PCAC,Llanes-Estrada_l1l2}.
Possibly a chiral quark model with the correct spin-tensor potentials will 
eventually reproduce the full spectrum of hadrons
\cite{Bicudo_thesis}.
Nevertheless this is only a quantitative problem, qualitatively the simple model 
used here is sufficient to study several implications of chiral symmetry and 
confinement.

Recently, the full mesonic spin-tensor potentials of the present simple model 
were determined for a quark and an antiquark with different isospin
\cite{Bicudo_last}. 
Here I exactly solve these boundstate equations of mesons and tachyons in different 
vacua. 
Importantly, the hamiltonian of this model can be approximately derived from QCD, 
\begin{eqnarray}
&&H=\int\, d^3x \left[ \psi^{\dag}( x) \;(m_0\beta -i{\vec{\alpha}
\cdot \vec{\nabla}} )\;\psi( x)\;+
{ 1\over 2} g^2 \int d^4y\, \
\right.
\nonumber \\
&&
\overline{\psi}( x)
\gamma^\mu{\lambda^a \over 2}\psi ( x)  
\langle A_\mu^a(x) A_\nu^b(y) \rangle
\;\overline{\psi}( y)
\gamma^\nu{\lambda^b \over 2}
 \psi( y)  \ + \ \cdots
\label{hamilt}
\end{eqnarray}
up to the first cumulant order, of two gluons
\cite{Bicudo_hvlt,Dosch,Kalashnikova,Nefediev},
which can be evaluated in the modified coordinate gauge,
\begin{equation}
g^2 \langle A_\mu^a(x) A_\nu^b(y) \rangle
\simeq-{3 \over 4} \delta_{ab} g_{\mu 0} 
g_{\nu 0}
\left[K_0^3({\bf x}-{\bf y})^2-U\right]
\label{potential}
\end{equation}
and this is a simple density-density harmonic effective 
confining interaction. $m_0$ is the current mass of the quark.
The infrared constant $U$ confines the quarks but the 
meson spectrum is completely insensitive to it. 
The important parameter is the potential strength $K_0$, 
the only physical scale in the interaction. 
In the true chiral symmetry breaking vacuum $K_0 \simeq 0.3\pm 0.05$ 
GeV fits reasonably the hadron spectra. However in the chiral
invariant vacuum the potential strength $K_0$ is supposed
to be greatly suppressed. For simplicity, I will consider
a vanishing light quark $m_0$ and all physical results will
scale only with the potential strength $K_0$.

I now address the meson and tachyon spectrum in different vacua.
In Section II the quark mass gap equation and the bound state 
quark-antiquark equation are reviewed. 
In Section III the mass gap and boundstate equations are solved 
numerically and the spectrum is studied in an arbitrary interpolation 
between the true and the chiral invariant vacuum. 
In Section IV the tachyons solutions of the boundstate equation
are analytically studied. 
These first studies are performed at vanishing
temperature. However in heavy ion collisions finite temperatures
are reached, sufficient for a QCD phase transition. The conclusion is 
presented in Section V, 
including the estimation of temperature effects on the spectra.

%
%
%TTTTTTTTTTTTTTTTTTTTTTTTTTTTTTTTTTTTTTTTTTTTTTTTTTTTTTTTTTT
%T                                          
%T                     TTTTTTTTTTT           11
%T                          T              1111
%T                          T                11
%T                          T                11
%T                          T                11
%T                          
%TTTTTTTTTTTTTTTTTTTTTTTTTTTTTTTTTTTTTTTTTTTTTTTTTTTTTTTTTTT
\begin{table}[t]
\caption{\label{algebraic} 
Matrix elements of the spin-dependent potentials
}
\begin{ruledtabular}
\begin{tabular}{c|ccccc}
$^{2S+1}L_J$					& $\delta_{{\bf S}_q,{\bf S}_{\bar q}}$ 
									& $ {\bf S}_q \hspace{-.075 cm} 
									\cdot \hspace{-.075 cm} {\bf S}_{\bar q}$ 
											& $({\bf S}_q + {\bf S}_{\bar q}) \hspace{-.075 cm} 
											\cdot \hspace{-.075 cm} {\bf L}$
													& $({\bf S}_q - {\bf S}_{\bar q})\hspace{-.075 cm} 
													\cdot \hspace{-.075 cm} {\bf L}$
														& tensor		\\
$^1S_0$ 						&1	&-3/4	&0		&0	&0 				\\
$^3P_0$ 						&1	&1/4	&-2		&0	&-1/3 			\\
$^3S_1$ 						&1	&1/4	&0		&0	&0 				\\
$^3D_1$ 						&1	&1/4	&-3		&0	&-1/6 			\\
$^3S_1\leftrightarrow {}^3D_1$	&0	&0		&0		&0	&$\sqrt{2}$/6 	\\
$^1P_1$ 						&1	&-3/4	&0		&0	&0 				\\
$^3P_1$ 						&1	&1/4	&-1		&0	&1/6 			\\
$^1P_1\leftrightarrow {}^3P_1$	&0	&0		&0		&$\sqrt{2}$	&0			 	\\
\end{tabular}
\end{ruledtabular}
\end{table}

\section{$T=0$ mass gap and boundstate equations}

\begin{figure}[t]
\caption{
The constituent quark masses $m_c(k)$, solutions of the mass gap equation,
for different current quark masses $m_0$. 
}\label{mass solution}
%figure created with latex and ghostview
\includegraphics[width=0.90\columnwidth]{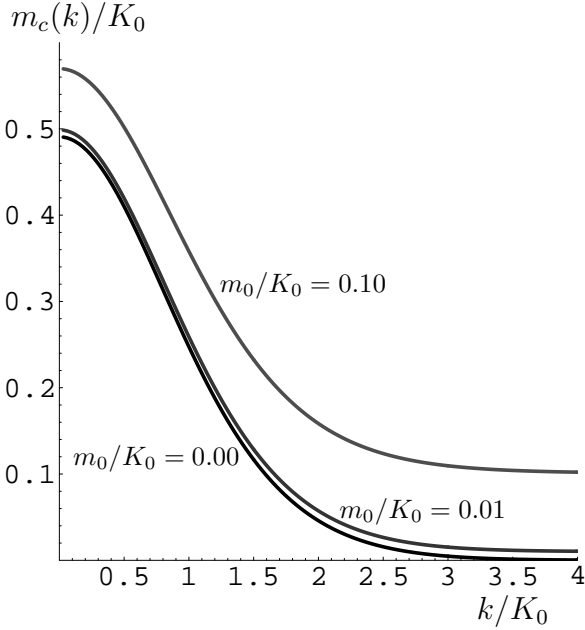}
\end{figure}

The relativistic invariant Dirac-Feynman propagators
\cite{Yaouanc}, 
can be decomposed in the quark and antiquark Bethe-Goldstone 
propagators
\cite{Bicudo_scapuz},
close to the formalism of non-relativistic quark models,
\FL
\begin{eqnarray}
{\cal S}_{Dirac}(k_0,\vec{k})
&=& {i \over \not k -m +i \epsilon}
\nonumber \\
&=& {i \over k_0 -E(k) +i \epsilon} \
\sum_su_su^{\dagger}_s \beta
\nonumber \\
&& - {i \over -k_0 -E(k) +i \epsilon} \
\sum_sv_sv^{\dagger}_s \beta \ ,
\nonumber \\
u_s({\bf k})&=& \left[
\sqrt{ 1+S \over 2} + \sqrt{1-S \over 2} \widehat k \cdot \vec \sigma \gamma_5
\right]u_s(0)  \ ,
\nonumber \\
v_s({\bf k})&=& \left[
\sqrt{ 1+S \over 2} - \sqrt{1-S \over 2} \widehat k \cdot \vec \sigma \gamma_5
\right]v_s(0)  \ ,
\nonumber \\
&=& -i \sigma_2 \gamma_5 u_s^*({\bf k}) \ ,
\label{propagators}
\end{eqnarray}
where $S=\sin(\varphi)={m_c\over \sqrt{k^2+m_c^2}} \ , 
\ C=\cos(\varphi)={k\over \sqrt{k^2+m_c^2}}$ and $\varphi$ is a chiral angle.
In the non condensed vacuum, $\varphi$ is equal to $\arctan{m_0 \over k}$,
but $\varphi$ is not determined from the onset when chiral symmetry breaking
occurs.
In the physical vacuum, the constituent quark mass $m_c(k)$, or the
chiral angle $\varphi(k)=\arctan{m_c(k) \over k}$, is a variational function
which is determined by the mass gap equation. Examples of solutions,
for different light current quark masses $m_0$, 
are depicted in Fig. \ref{mass solution}.
For simplicity in the remaining iof this paper $m_0=0$ will be assumed,
nevertheless the effect of a finite current quark mass can be estimated
with a small increase of the dynamically generated constituent quark mass
$m_c$.

%
%%%%%%%%%%%%%%%%%%%%%%%%%%%%%%%%%%%%%%%
%%                                   %%
%%                FFFFFF             %%
%%                F                  %%
%%                FFF                %%
%%                F                  %%
%%                F                  %%
%%                                   %%
%%%%%%%%%%%%%%%%%%%%%%%%%%%%%%%%%%%%%%%

Then there are three equivalent methods to find 
the true and stable vacuum, where constituent quarks acquire 
the constituent mass.
One method consists in assuming a quark-antiquark $^3P_0$ 
condensed vacuum, and in minimizing the vacuum energy density. 
A second method consists in rotating the quark and antiquark 
fields with a Bogoliubov-Valatin canonical transformation 
to diagonalize the terms in the hamiltonian with two   
quark or antiquark second quantized fields. 
A third method consists in solving the Schwinger-Dyson 
equations for the propagators. Any of these methods
lead to the same mass gap equation and to the quark 
dispersion relation. Here I replace the propagator
of eq. (\ref{propagators}) in the Schwinger-Dyson equation, 
\begin{eqnarray}
\label{2 eqs}
&&0 = u_s^\dagger(k) \left\{k \widehat k \cdot \vec \alpha + m_0 \beta
-\int {d w' \over 2 \pi} {d^3k' \over (2\pi)^3}
i V(k-k') \right.
\nonumber \\
&&\left. \sum_{s'} \left[ { u(k')_{s'}u^{\dagger}(k')_{s'} 
 \over w'-E(k') +i\epsilon}
-{ v(k')_{s'}v^{\dagger}(k')_{s'} 
  \over -w'-E(k')+i\epsilon} \right]
\right\} v_{s''}(k) \  \
\nonumber \\
&&E(k) = u_s^\dagger(k) \left\{k \widehat k \cdot \vec \alpha + m_0 \beta
-\int {d w' \over 2 \pi} {d^3k' \over (2\pi)^3}
i V(k-k')  \right.
\nonumber \\
&&\left. \sum_{s'} \left[ { u(k')_{s'}u^{\dagger}(k')_{s'} 
 \over w'-E(k') +i\epsilon}
-{   v(k')_{s'}v^{\dagger}(k')_{s'}  
 \over -w'-E(k')+i\epsilon} \right]
\right\} u_s(k),
\end{eqnarray}
where, with the simple density-density harmonic interaction
\cite{Yaouanc}, the integral of the potential is a laplacian 
and the mass gap equation and the quark energy are finally,
\begin{eqnarray}
\label{mass gap}
\Delta \varphi(k)  &=& 2 k S(k) -2 m_0 C(k) - { 2 S(k) C(k) \over k^2 }  
\\ \nonumber 
E(k)&=& k C(k) + m_0 S(k) - { {\varphi'(k) }^2 \over 2 } - { C(k)^2 \over k^2 } 
+{ U \over 2} \ .
\end{eqnarray}
Numerically, this equation is a non-linear ordinary differential
equation. It can be solved with the Runge-Kutta and shooting method.
Examples of solutions for the current quark mass $m_c(k)= k \tan \varphi$, 
for different current quark masses $m_0$, 
are depicted in Fig. \ref{mass solution}.

%
%
%TTTTTTTTTTTTTTTTTTTTTTTTTTTTTTTTTTTTTTTTTTTTTTTTTTTTTTTTTTT
%T                                          
%T                     TTTTTTTTTTT           11
%T                          T              1111
%T                          T                11
%T                          T                11
%T                          T                11
%T                          
%TTTTTTTTTTTTTTTTTTTTTTTTTTTTTTTTTTTTTTTTTTTTTTTTTTTTTTTTTTT
\begin{table}[t]
\caption{\label{spin dependent} 
The positive and negative energy spin-independent, spin-spin, spin-orbit and 
tensor potentials are shown, for the simple density-density harmonic model
of eq. (\ref{potential}).
$\varphi'(k)$, ${\cal C}(k)$ and ${\cal G}(k)= 1 - S(k) $ are all functions of the constituent 
quark(antiquark) mass.
}
\begin{ruledtabular}
\begin{tabular}{c|c}
& $V^{++}=V^{--}$  \\ \hline
spin-indep. & $- {d^2 \over dk^2 } + { {\bf L}^2 \over k^2 } + 
{1 \over 4} \left( {\varphi'_q}^2 + {\varphi'_{\bar q}}^2 \right) 
+ {  1 \over k^2} \left( {\cal G}_q +{\cal G}_{\bar q}  \right) -U $  \\ 
spin-spin & $ {4 \over 3 k^2} {\cal G}_q {\cal G}_{\bar q} {\bf S}_q \cdot {\bf S}_{\bar q} $  \\ 
spin-orbit & $ {1 \over  k^2} \left[ \left( {\cal G}_q + 
{\cal G}_{\bar q} \right) \left( {\bf S}_q +{\bf S}_{\bar q}\right) 
+\left( {\cal G}_q - {\cal G}_{\bar q} \right) \left( {\bf S}_q -{\bf S}_{\bar q}\right)  \right]
\cdot {\bf L} $  \\ 
tensor & $ -{2 \over  k^2} {\cal G}_q {\cal G}_{\bar q} 
\left[ ({\bf S}_q \cdot \hat k ) ({\bf S}_{\bar q} \cdot \hat k )
-{1 \over 3} {\bf S}_q \cdot {\bf S}_{\bar q} \right] $ \\ \hline
& $V^{+-}=V^{-+}$ \\ \hline
spin-indep. & $0$  \\ 
spin-spin & $ -{4 \over 3} \left[ {1\over 2} {\varphi'_q} {\varphi'_{\bar q}} + 
{1\over k^2} {\cal C}_q {\cal C}_{\bar q}  \right]
{\bf S}_q \cdot {\bf S}_{\bar q} $  \\ 
spin-orbit & $0$  \\ 
tensor & $ \left[ -2 {\varphi'_q} {\varphi'_{\bar q}} + 
{2\over k^2} {\cal C}_q {\cal C}_{\bar q}  \right]
\left[ ({\bf S}_q \cdot \hat k ) ({\bf S}_{\bar q} \cdot \hat k )
-{1 \over 3} {\bf S}_q \cdot {\bf S}_{\bar q} \right] $
\end{tabular}
\end{ruledtabular}
\end{table}

The Salpeter-RPA equations for a meson (a colour singlet
quark-antiquark bound state) can be derived from the Lippman-Schwinger
equations for a quark and an antiquark, or replacing the propagator
of eq. (\ref{propagators}) in the Bethe-Salpeter equation. In either way, one gets
\cite{Bicudo_scapuz}
\FL
\begin{eqnarray}
\label{homo sal}
\phi^+(k,P) &=& { u^\dagger(k_1) \chi(k,P)  v(k_2) 
\over +M(P)-E(k_1)-E(k_2) }
\nonumber \\
{\phi^-}^t(k,P) &=& { v^\dagger(k_1) \chi(k,P) u(k_2)
\over -M(P)-E(k_1)-E(k_2)}
\nonumber \\
\chi(k,P) &=&
\int {d^3k' \over (2\pi)^3} V(k-k') \left[ 
u(k'_1)\phi^+(k',P)v^\dagger(k'_2) \right.
\nonumber \\
&&\left. +v(k'_1){\phi^-}^t(k',P) u^\dagger(k'_2)\right] 
\end{eqnarray}
where $k_1=k+{P \over 2} \ , \ k_2=k-{P \over 2}$ and $P$ is
the total momentum of the meson.
Notice that, solving for $\chi$, one gets the Salpeter equations of Yaouanc et al.
\cite{Yaouanc}.

The Salpeter-RPA equations of PB et al. 
\cite{Bicudo_thesis}
and of Llanes-Estrada et al. 
\cite{Llanes-Estrada_thesis}
are obtained deriving the equation for the positive energy wavefunction
$\phi^+$ and for the negative energy wavefunction $\phi^-$. The 
relativistic equal time equations have the double of coupled
equations than the Schr\"odinger equation, although in many cases the
negative energy components can be quite small. This results in four potentials 
$V^{\alpha \beta}$ respectively coupling $\nu^\alpha=r \phi^\alpha$ to 
$\nu^\beta$. The Pauli $\vec \sigma$ matrices in the spinors of eq. (\ref{propagators}) 
produce the spin-dependent
\cite{Bicudo_baryon} 
potentials of Table \ref{spin dependent}. 

Notice that both the pseudoscalar and scalar equations
have a system with two equations. This is the minimal number of relativistic 
equal time equations. However the spin-dependent interactions 
couple an extra pair of equations both in the vector and axialvector channels.
While the coupling of the s-wave and the d-wave are standard in vectors, the coupling 
of the spin-singlet and spin-triplet in axialvectors only occurs if the quark and antiquark 
masses are different, say in heavy-light systems. 
I now combine the algebraic matrix elements of Table \ref{algebraic}
with the spin-dependent potentials of Table \ref{spin dependent},
to derive the full Salpeter-RPA radial boundstate
equations (where the infrared $U$ is dropped from now on). 
I get the $J^P=0^{-}$,  $^1 S_0$ pseudoscalar ($P$) equations,

\onecolumngrid
\begin{equation}
\label{pseudoscalar}
\left\{ \left( -{d^2 \over d k ^2} +E_q(k) +E_{\bar q}(k)  
+ { {\varphi'_q}^2 +{\varphi'_{\bar q}}^2 \over 4} + {1-S_q S_{ \bar q} \over k^2}  \right)
\left[ \begin{array}{cc}
1 & 0 \\ 0 & 1 \end{array} \right]
+
\left( {\varphi'_q \varphi'_{\bar q} \over 2} + {C_q C_{\bar q} \over k^2 } \right)
\left[ \begin{array}{cc}
0 & 1 \\ 1 & 0 \end{array} \right]
-M 
\left[ \begin{array}{cc}
1 & 0 \\ 0 & -1 \end{array} \right]
\right\}
\left( \begin{array}{c} \nu_{^1S_0}^+(k) \\ \nu_{^1S_0}^-(k) \end{array} \right) = 0
\  ,
\end{equation}
the $J^P=0^{+}$, $^3 P_0$ scalar ($S$) equations,
\begin{equation}
\label{scalar}
\left\{ \left( -{d^2 \over d k ^2} +E_q(k) +E_{\bar q}(k)  
+ { {\varphi'_q}^2 +{\varphi'_{\bar q}}^2 \over 4} + {1+S_q S_{ \bar q} \over k^2}  \right)
\left[ \begin{array}{cc}
1 & 0 \\ 0 & 1 \end{array} \right]
+
\left( {\varphi'_q \varphi'_{\bar q} \over 2} - {C_q C_{\bar q} \over k^2 } \right)
\left[ \begin{array}{cc}
0 & 1 \\ 1 & 0 \end{array} \right]
-M 
\left[ \begin{array}{cc}
1 & 0 \\ 0 & -1 \end{array} \right]
\right\}
\left( \begin{array}{c} \nu_{^3P_0}^+(k) \\ \nu_{^3P_0}^-(k) \end{array} \right) = 0
\  .
\end{equation}
the $J^P=1^{-}$,  coupled $^3 S_1$ and $^3D_1$ vector ($V$ and $V^*$) equations ,
\begin{eqnarray}
\label{vector}
\left\{ 
\left( -{d^2 \over d k ^2} +E_q(k) +E_{\bar q}(k)  
+ { {\varphi'_q}^2 +{\varphi'_{\bar q}}^2 \over 4} 
+ {7-4S_q -4S_{ \bar q}+S_q S_{ \bar q}\over 3 k^2} \right)
\left[ \begin{array}{cccc}
1 & 0 & 0 & 0 \\ 
0 & 1 & 0 & 0 \\ 
0 & 0 & 0 & 0 \\ 
0 & 0 & 0 & 0 
\end{array} \right]
+
\left( -{\varphi'_q \varphi'_{\bar q} \over 6} 
-{C_q C_{\bar q} \over 3k^2 } \right)
\left[ \begin{array}{cccc}
0 & 1 & 0 & 0 \\ 
1 & 0 & 0 & 0 \\
0 & 0 & 0 & 0 \\
0 & 0 & 0 & 0
\end{array} \right]
\right. &&
\\ \nonumber 
+
\left( -{d^2 \over d k ^2} +E_q(k) +E_{\bar q}(k)  
+ { {\varphi'_q}^2 +{\varphi'_{\bar q}}^2 \over 4} 
+ {8+4S_q +4S_{ \bar q}+2S_q S_{ \bar q}\over 3 k^2} \right)
\left[ \begin{array}{cccc}
0 & 0 & 0 & 0 \\ 

0 & 0 & 0 & 0 \\ 
0 & 0 & 1 & 0 \\ 
0 & 0 & 0 & 1 
\end{array} \right]
+
\left(  {\varphi'_q \varphi'_{\bar q} \over 6} 
-{ 2 C_q C_{\bar q} \over 3k^2 } \right)
\left[ \begin{array}{cccc}
0 & 0 & 0 & 0 \\ 
0 & 0 & 0 & 0 \\
0 & 0 & 0 & 1 \\
0 & 0 & 1 & 0
\end{array} \right] 
&&
\\ \nonumber 
\left.
- 
{ \left(1-S_q\right) \left(1-S_{ \bar q}\right)\over 3 k^2 } 
\left[ \begin{array}{cccc}
0 & 0 & \sqrt{2} & 0 \\ 
0 & 0 & 0 & \sqrt{2} \\ 
\sqrt{2} & 0 & 0 & 0 \\ 
0 & \sqrt{2} & 0 & 0 

\end{array} \right]
-\left(  {\varphi'_q \varphi'_{\bar q} \over 3} 
- {C_q C_{\bar q} \over 3k^2 } \right) 
\left[ \begin{array}{cccc}
0 & 0 & 0 & \sqrt{2} \\ 
0 & 0 & \sqrt{2} & 0 \\
0 & \sqrt{2} & 0 & 0 \\
\sqrt{2} & 0 & 0 & 0
\end{array} \right]
-M 
\left[ \begin{array}{cccc}
1 & 0 & 0 & 0 \\ 
0 & -1 & 0 & 0 \\ 
0 & 0 & 1 & 0 \\ 
0 & 0 & 0 & -1 
\end{array} \right]
\right\} 
\left( \begin{array}{c} \nu_{^3S_1}^+(k) \\ \nu_{^3S_1}^-(k) \\ \nu_{^3D_1}^+(k) \\ \nu_{^3D_1}^-(k) 
\end{array} \right) 
& =& 0
\  ,
\end{eqnarray}
the $J^P=1^{+}$, coupled $^1P_1$ and $^3P_1$ axialvector ($A$ and $A^*$) equations 
\begin{eqnarray}
\label{axialvector}
\left\{ \left( -{d^2 \over d k ^2} +E_q(k) +E_{\bar q}(k)  
+ { {\varphi'_q}^2 +{\varphi'_{\bar q}}^2 \over 4} 
+ {3-S_q S_{ \bar q} \over k^2}  \right)
\left[ \begin{array}{cccc}
1 & 0 & 0 & 0 \\ 
0 & 1 & 0 & 0 \\ 
0 & 0 & 0 & 0 \\ 
0 & 0 & 0 & 0 
\end{array} \right]
+
\left( {\varphi'_q \varphi'_{\bar q} \over 2} + {C_q C_{\bar q} \over k^2 } \right)
\left[ \begin{array}{cccc}
0 & 1 & 0 & 0 \\ 
1 & 0 & 0 & 0 \\
0 & 0 & 0 & 0 \\
0 & 0 & 0 & 0
\end{array} \right]
\right.
&&
\\ \nonumber 
\left( -{d^2 \over d k ^2} +E_q(k) +E_{\bar q}(k)  
+ { {\varphi'_q}^2 +{\varphi'_{\bar q}}^2 \over 4} 
+{2 \over k^2} \right)
\left[ \begin{array}{cccc}
0 & 0 & 0 & 0 \\ 
0 & 0 & 0 & 0 \\ 
0 & 0 & 1 & 0 \\ 
0 & 0 & 0 & 1 
\end{array} \right]
+
\left( - {\varphi'_q \varphi'_{\bar q} \over 2} \right)
\left[ \begin{array}{cccc}
0 & 0 & 0 & 0 \\ 
0 & 0 & 0 & 0 \\
0 & 0 & 0 & 1 \\
0 & 0 & 1 & 0
\end{array} \right] 
&&
\\ \nonumber 
\left.
+
{ S_q - S_{\bar q}\over  k^2 } 
\left[ \begin{array}{cccc}
0 & 0 & \sqrt{2} & 0 \\ 
0 & 0 & 0 & \sqrt{2} \\ 
\sqrt{2} & 0 & 0 & 0 \\ 
0 & \sqrt{2} & 0 & 0 
\end{array} \right]
-M 
\left[ \begin{array}{cccc}
1 & 0 & 0 & 0 \\ 
0 & -1 & 0 & 0 \\ 
0 & 0 & 1 & 0 \\ 
0 & 0 & 0 & -1 
\end{array} \right]
\right\} 
\left( \begin{array}{c} \nu_{^1P_1}^+(k) \\ \nu_{^1P_1}^-(k) \\ \nu_{^3P_1}^+(k) \\ \nu_{^3P_1}^-(k) 
\end{array} \right) 
& =& 0
\  ,
\end{eqnarray}

\twocolumngrid

\section{Numerical solution of the mass gap and boundstate equations at $T=0$}

In the light-light limit of $m_q =m_{\bar q} \rightarrow 0$ and $\varphi \rightarrow 0$, 
it is clear that eq. (\ref{pseudoscalar}) and eq. (\ref{scalar}) become identical. They
also possess takyonic solutions 
\cite{Yaouanc}. 
In the same limit, eq. (\ref{vector}) can be block diagonalized
\cite{Yaouanc}, 
and each block, with mixed s-wave and d-wave, is identical one of the two 
independent blocks of eq. (\ref{axialvector}). This checks that the chiral partners 
$P$-$S$ and $V,V^*$-$A,A^*$ are degenerate in the false chiral symmetric vacuum.

Another interesting case is the heavy-light case where, say, the antiquark has a mass 
$m_{\bar q}\simeq {m_0}_{\bar q} >> K_0$, there are no Tachyons, and the negative energy 
components nearly vanish, like in non-relativistic quark models. 
In the infinite $m_{\bar q}$ limit, $S_{\bar q} \rightarrow 1$, and the antiquark spin 
is irrelevant, see Table \ref{spin dependent}, complying with the 
Isgur-Wise heavy-quark symmetry.

%
%%%%%%%%%%%%%%%%%%%%%%%%%%%%%%%%%%%%%%%
%%                                   %%
%%                FFFFFF             %%
%%                F                  %%
%%                FFF                %%
%%                F                  %%
%%                F                  %%
%%                                   %%
%%%%%%%%%%%%%%%%%%%%%%%%%%%%%%%%%%%%%%%
\begin{figure}[t]
\caption{
Light-light meson masses,
in (a), pseudoscalar,
in (b), scalar,
when the light quark mass interpolates from the zero mass of 
the chiral invariant false vacuum to the solution $m_c$ of the 
mass gap equation in the true vacuum. The dark curves correspond
to mesonic real masses and the light curves correspond
to tachyonic imaginary masses.
}\label{light-light PS}
%figure created with latex and ghostview
\includegraphics[width=1.00\columnwidth]{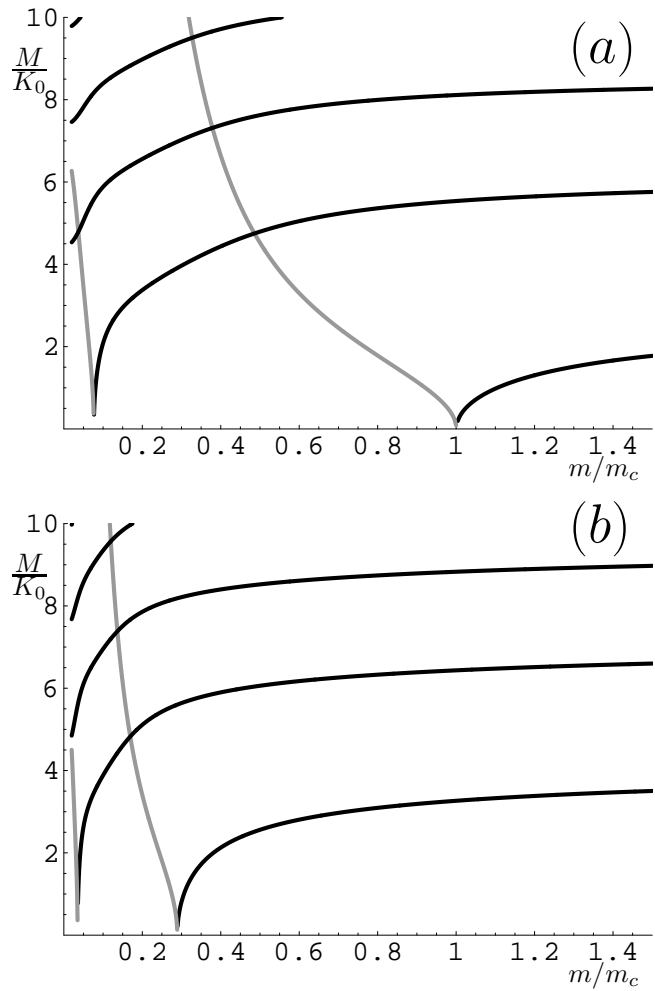}
\end{figure}

%
%%%%%%%%%%%%%%%%%%%%%%%%%%%%%%%%%%%%%%%
%%                                   %%
%%                FFFFFF             %%
%%                F                  %%
%%                FFF                %%
%%                F                  %%
%%                F                  %%
%%                                   %%
%%%%%%%%%%%%%%%%%%%%%%%%%%%%%%%%%%%%%%%
\begin{figure}[t]
\caption{
Light-light meson masses,
in (a), vector and
in (b), axial,
when the light quark mass interpolates from the zero mass of 
the chiral invariant false vacuum to the solution $m_c$ of the 
mass gap equation in the true vacuum. The dark curves correspond
to mesonic real masses and the light curves correspond
to tachyonic imaginary masses.
}\label{light-light VA}
%figure created with latex and ghostview
\includegraphics[width=1.00\columnwidth]{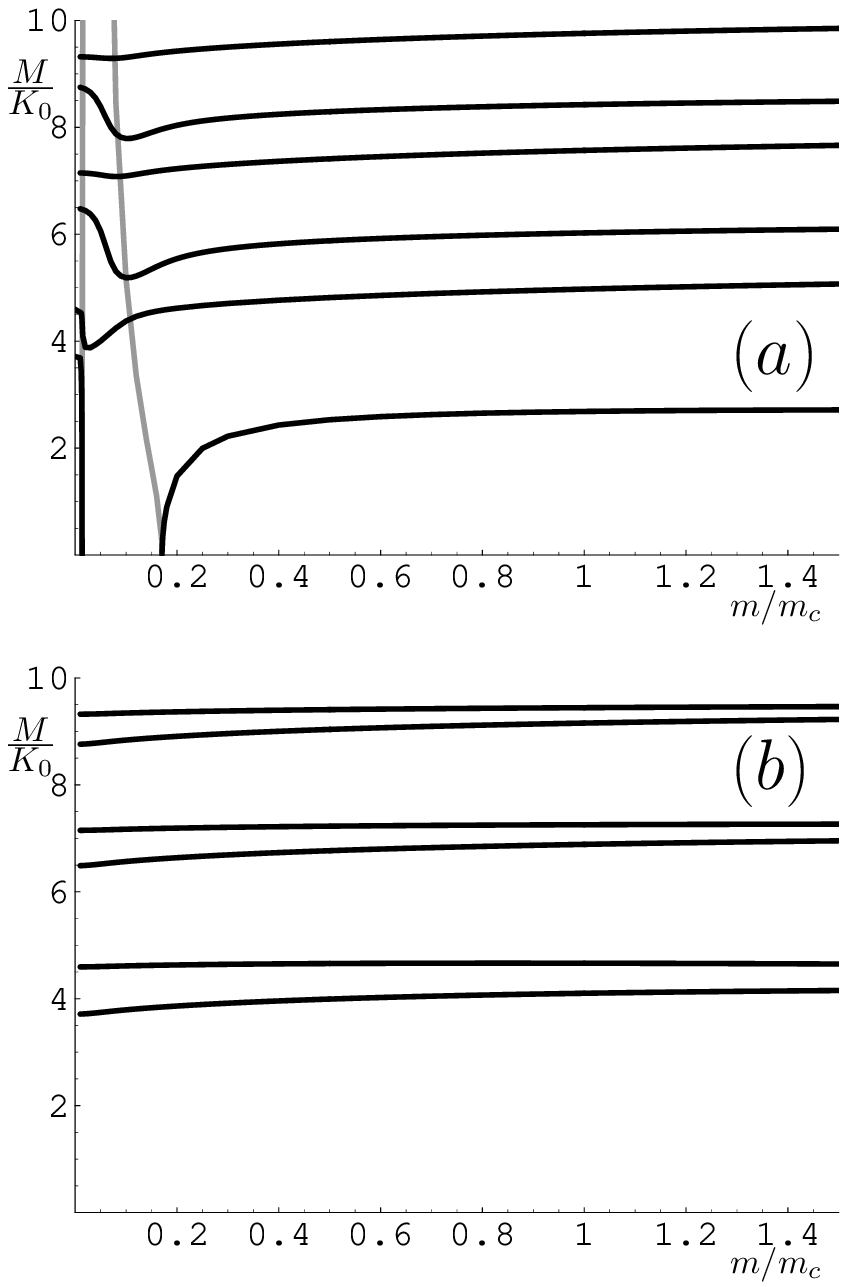}
\end{figure}

%
%%%%%%%%%%%%%%%%%%%%%%%%%%%%%%%%%%%%%%%
%%                                   %%
%%                FFFFFF             %%
%%                F                  %%
%%                FFF                %%
%%                F                  %%
%%                F                  %%
%%                                   %%
%%%%%%%%%%%%%%%%%%%%%%%%%%%%%%%%%%%%%%%
\begin{figure}[t]
\caption{
Heavy-light meson masses minus the infinitely heavy antiquark
mass,
in (a), pseudoscalar,
in (b), scalar,
when the light quark mass interpolates from the zero mass of 
the chiral invariant false vacuum to the solution $m_c$ of the 
mass gap equation in the true vacuum. In this case there
are no tachyonic imaginary masses.
}\label{heavy-light PS}
%figure created with latex and ghostview
\includegraphics[width=1.00\columnwidth]{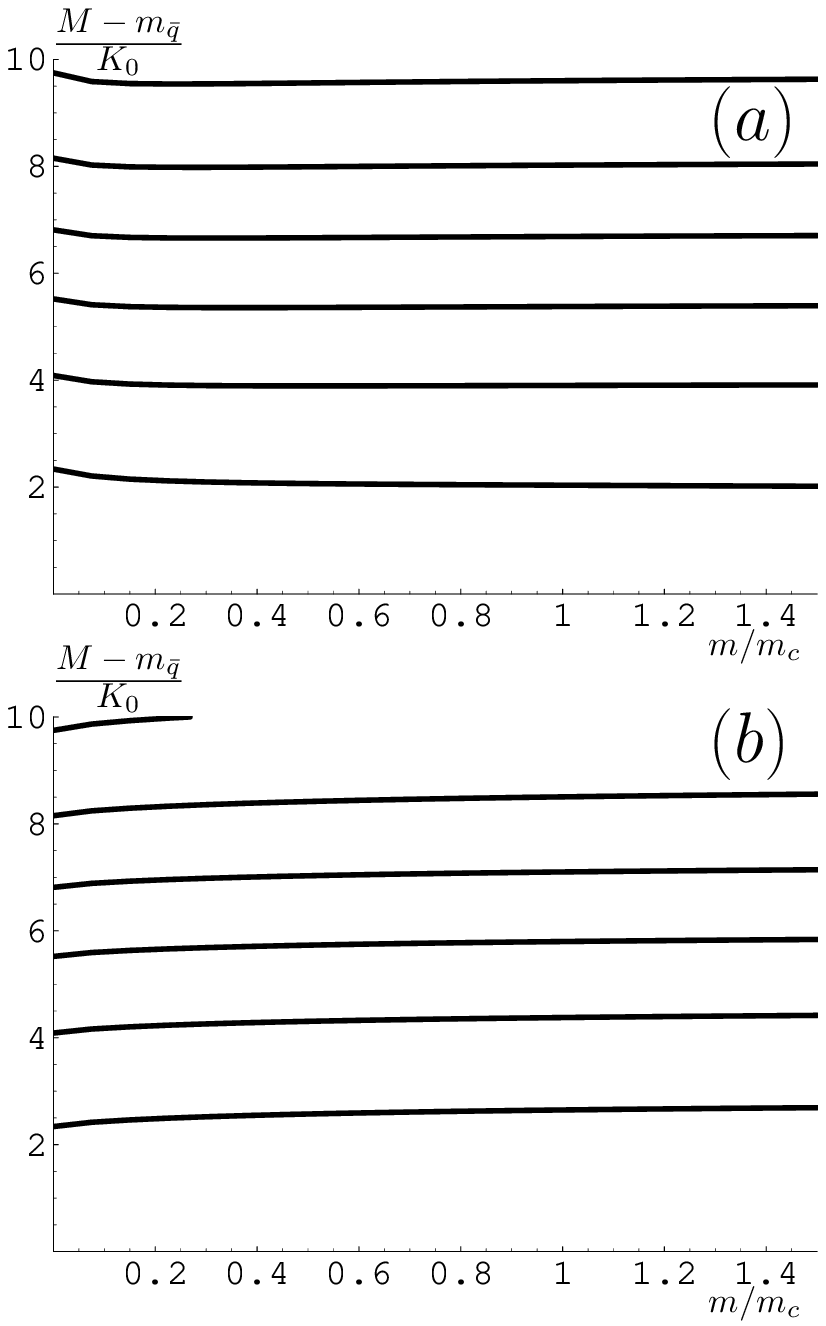}
\end{figure}

%
%%%%%%%%%%%%%%%%%%%%%%%%%%%%%%%%%%%%%%%
%%                                   %%
%%                FFFFFF             %%
%%                F                  %%
%%                FFF                %%
%%                F                  %%
%%                F                  %%
%%                                   %%
%%%%%%%%%%%%%%%%%%%%%%%%%%%%%%%%%%%%%%%
\begin{figure}[t]
\caption{
Heavy-light meson masses minus the infinitely heavy antiquark
mass,
in (a), vector and
in (b), axial,
when the light quark mass interpolates from the zero mass of 
the chiral invariant false vacuum to the solution $m_c$ of the 
mass gap equation in the true vacuum. In this case there
are no tachyonic imaginary masses.
}\label{heavy-light VA}
%figure created with latex and ghostview
\includegraphics[width=1.00\columnwidth]{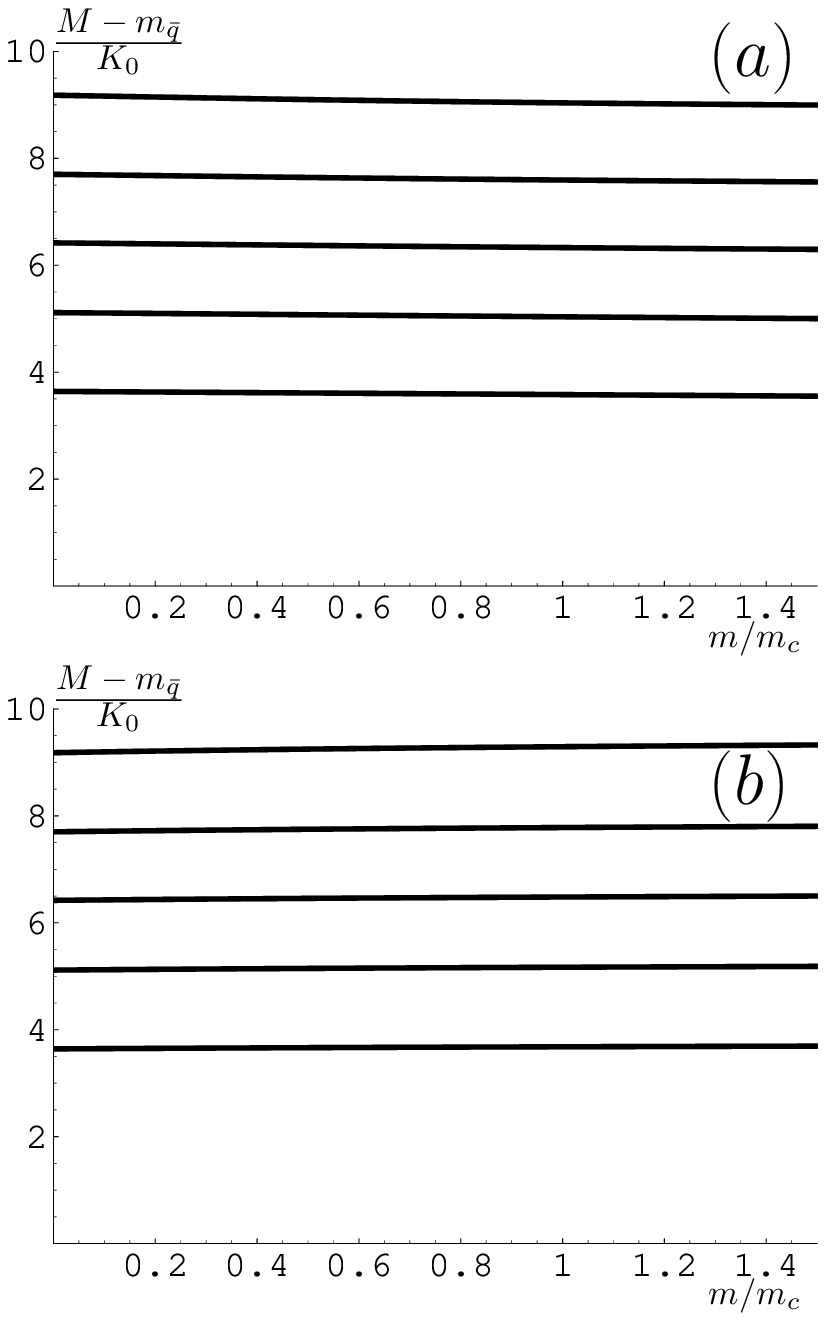}
\end{figure}

Notice that this model, like any chiral model, has the same number 
of meson states in the spectrum as  the normal quark model. The mass 
splittings can de related, as usual, to spin-tensor potentials.

For the numerical solution, I change the sign of the 
second and fourth lines in eqs (\ref{pseudoscalar}) to 
(\ref{axialvector}) and then I get a simple eigenvalue 
equation. The results are shown in Figs. \ref{light-light PS},
\ref{light-light VA}, \ref{heavy-light PS} and \ref{heavy-light VA}. 
In Fig. \ref{light-light PS}, the pseudoscalar and scalar
light quark and light antiquark meson
masses are interpolated from the true spontaneously chiral
symmetry breaking vacuum to the false chiral restored vacuum.
In Fig. \ref{light-light VA}, the vector and axial
light quark and light antiquark meson
masses are interpolated from the true spontaneously chiral
symmetry breaking vacuum to the false chiral restored vacuum.
In Fig. \ref{heavy-light PS}, the pseudoscalar and scalar
light quark and heavy antiquark meson
masses are interpolated from the true spontaneously chiral
symmetry breaking vacuum to the false chiral restored vacuum.
In Fig. \ref{heavy-light VA}, the vector and axial
light quark and heavy antiquark meson
masses are interpolated from the true spontaneously chiral
symmetry breaking vacuum to the false chiral restored vacuum.

A remarkable result of the numerical finite difference solutions 
is that all studied pseudoscalar and scalar mesons, including 
all radial excitations, become tachyons, with arbitrarily
large imaginary masses. This will be confirmed in the next
Section IV. 

On the other hand, all the other mesons suffer small mass over 
potential strength $M/K_0$ corrections from one vacuum to the other. 
Notice however that the potential strength $K_0$ is expected to 
change significantly when the vacuum is changed.
This will be discussed in detail in Section V.

A remarkable feature of the vector meson groundstate in
Fig. \ref{light-light PS} may be relevant for the $\rho$ 
and $\omega$ mesons. Although the $M/K_0$ corrections from 
one vacuum to the other are small, at small but non-vanishing 
quark mass the groundstate vector meson is a tachyon. This
occurs just before the vector and axial vector are degenerate.
Because the actual light current quark masses are small but 
non-vanishing, this will be addressed in Section V

\section{Analytical study of the Tachyons }

I now study in detail the properties of the eigenvalues
of the Salpeter or RPA equations. The boundstate equation
can be decoupled,
\begin{eqnarray}
&&
\left\{
\begin{array}{ccc}
H \nu^+ + & V \nu^- =& M \nu^+ \\
H \nu^+ + & V \nu^- =& - M \nu^- 
\end{array}
\right.
\nonumber \\
&\Rightarrow&
\left\{
\begin{array}{cc}
( H + V) (\nu^+ +  \nu^-) =& M (\nu^+ -  \nu^-) \\
( H - V) (\nu^+ -  \nu^-) =& M (\nu^+ +  \nu^-) \\
\end{array}
\right.
\\ \nonumber 
&\Rightarrow&
\left\{
\begin{array}{cc}
( H - V) ( H + V) (\nu^+ +  \nu^-) =& M^2 (\nu^+ +  \nu^-) \\
( H + V) ( H - V) (\nu^+ -  \nu^-) =& M^2 (\nu^+ -  \nu^-) \\
\end{array}
\right.
\ .
\label{decoupled}
\end{eqnarray}
Thus we get a pair of eigenvalue equations, where $H$ and $V$ are hermitean,
but $( H - V) ( H + V)$ and $( H + V) ( H - V)$ are not hermitean. Nevertheless
$M^2$ can be prooved to be real. if one considers two different eigenvalues 
$M_1{}^2$ and $M_2{}^2$, 
\begin{eqnarray}
&&
\left\{
\begin{array}{cc}
( H - V) ( H + V) (\nu_1{}^+ +  \nu_1{}^-) =& M_1{}^2 (\nu_1{}^+ +  \nu_1{}^-) \\
(\nu_2{}^+ -  \nu_2{}^-)^\dagger( H - V) ( H + V) =& (\nu_2{}^+ -  \nu_2{}^-)^\dagger {M_2{}^2}^* \\
\end{array}
\right.
\nonumber \\
&&\Rightarrow
\left\{
\begin{array}{c}
( M_1{}^2 -{M_2{}^2}^* )(\nu_2{}^+ -  \nu_2{}^-)^\dagger (\nu_1{}^+ +  \nu_1{}^-) = 0 \\
( M_2{}^2 -{M_1{}^2}^* )(\nu_1{}^+ -  \nu_1{}^-)^\dagger (\nu_2{}^+ +  \nu_2{}^-) = 0 \\
\end{array}
\right. \ .
\end{eqnarray}
Notice that the orthonormalization condition
\cite{Yaouanc,Bicudo_thesis,Bicudo_scapuz}
of the Salpeter-RPA equation is,
\begin{equation}
\left( {\nu_i{}^+ }^\dagger \, , \, {\nu_i{}^-}^\dagger \right)
\left[ \begin{array}{cc}
1 & 0 \\
0 & -1\\
\end{array} \right]
\left( \begin{array}{c}
\nu_j{}^+ \\
\nu_j{}^- \\
\end{array} \right)
= \delta_{i,j} \ \  .
\end{equation}
Thus, either the two eigenvectors are orthogonal or the squared eigenvalue $M^2$ is real.
This shows that the solutions of the boundstate equation can only have 
real or purely imaginary masses. While the real masses correspond to mesons, the imaginary 
masses correspond to tachyons.

I now study in detail the solutions in the chiral invariant vacuum and in the chiral limit,
where both the current mass $m_0$ and the constituent mass $m_c$ vanish. 
In general the boundstate equations decouple in two different equations, one for 
$J \geq 0$ with,
\begin{equation}
\label{J g 0}
\left\{ 
\begin{array}{cl}
H=& -{d^2 \over d k ^2} +2 k  - {1\over k^2} + {j(j+1) \over k^2} \\ 
V=& {1\over k^2} \\
\end{array} \right. \ ,
\end{equation}
and another for $J \geq 1$ with,
\begin{equation}
\label{J g 1}
\left\{ 
\begin{array}{cl}
H=& -{d^2 \over d k ^2} +2 k  - {2\over k^2} + {j(j+1) \over k^2}\\ 
V=& {0\over k^2} \\
\end{array} \right. \ .
\end{equation}
Notice that the different potentials 
$ -{d^2 \over d k ^2} \, , \
2 k \, , \ {1\over k^2} $ are bound from below and 
positive definite in the sense that all their eigenvalues
are positive. However $-{1\over k^2} $ is unbound from
below. Thus, in eq. (\ref{decoupled}) all terms $H+V$ or
$H-V$ are positive definite and bound from below, except
for the $H-V$ of the $J=0$ pseudoscalar and scalar tachyons in
eq. (\ref{J g 0}). 

Notice that Fig. \ref{light-light PS} suggests that all
pseudoscalars and scalars become tachyons in the chiral
invariant vacuum. To confirm this suggestion of an infinite 
number of tachyons it is convenient to regularize the scalar 
and pseudoscalar equations, because the wave-functions 
are concentrated at extremely small distances. 
A very small quark mass $m$ is assumed constant
for simplicity, and the momentum and mass are rescaled,
\begin{eqnarray}
k / m &\rightarrow& {k'} \ ,
\nonumber \\
M \, m^2 &\rightarrow& M' \ .
\end{eqnarray}
Notice that any finite solution $M'$ in fact 
corresponds to an infinite mass $M=M' /m^2 $, and that 
a wave-function with a finite $k'$ corresponds 
to a wave-function with infinitesimal momentum $k=k'm$.
Then, starting from eq. (\ref{pseudoscalar}) 
one gets for the pseudoscalar,
\begin{eqnarray}
\left\{ 
\begin{array}{cl}
H+V=& -{d^2 \over d {k'}^2}    \\ 
H-V=& -{d^2 \over d {k'}^2}  -{2\over k^2+1} -{1\over (k^2+1)^2} \\
\end{array} \right. \ ,
\end{eqnarray}
respectively positive definite and with negative eigenvalues,
and from eq. (\ref{scalar}) one gets for the scalar,
\begin{eqnarray}
\left\{ 
\begin{array}{cl}
H-V=& -{d^2 \over d {k'} ^2}  +{2\over {k'}^2({k'}^2+1)} -{1\over ({k'}^2+1)^2} \\ 
H+V=& -{d^2 \over d {k'} ^2}  +{2\over {k'}^2({k'}^2+1)} -{2\over {k'}^2+1} \\
\end{array} \right. \ ,
\end{eqnarray}
respectively positive definite and with negative
eigenvalues. An irrelevant term $m^3 {2 {k'} / \sqrt{{k'}^2+1}}$ is also present
in the rescaled equations. 

The Bohr-Sommerfeld quantization condition can be used to count the 
number of negative eigenvalues of the $H-V$ pseudoscalar operator 
and of the $H+V$ scalar operator. The leading term at high momentum, 
assuming the highest possible negative mass $M' \simeq 0$, is,
\begin{equation}
\int_0^\infty \sqrt{1 \over 1 + {k'}^2} \, d{k'}= \infty \ .
\end{equation}
This shows that the number of tachyons in the pseudoscalar and scalar
channels are both infinite.

This is confirmed by the numerical solution of the regularized Salpeter equation. In Table 
\ref{tachyon+meson} we show the masses of the different light-light tachyons and mesons in 
the chiral invariant false vacuum and in the chiral limit.

Notice that there is no pseudoscalar-scalar
degeneracy in the rescaled equations since the equations are different, ${M'}_{S} \neq {M'}_{PS}$. 
Nevertheless both pseudoscalar and scalar tachyons have infinite imaginary masses and we get
$M_S = M_P= \infty \, i $.

%
%
%TTTTTTTTTTTTTTTTTTTTTTTTTTTTTTTTTTTTTTTTTTTTTTTTTTTTTTTTTTT
%T                                          
%T                     TTTTTTTTTTT           11
%T                          T              1111
%T                          T                11
%T                          T                11
%T                          T                11
%T                          
%TTTTTTTTTTTTTTTTTTTTTTTTTTTTTTTTTTTTTTTTTTTTTTTTTTTTTTTTTTT
\begin{table}[t]
\caption{\label{tachyon+meson} 
Masses of the first angular and radial excitations of the different 
light-light tachyons and mesons in the chiral invariant false vacuum and 
in the chiral limit. Each column includes both positive and 
negative parity degenerate states, except for the pseudoscalar 
and scalar tachyonic states. Notice that the tachyon masses are 
infinite and that they are regularized by an arbitrarily small 
quark mass $m$. The meson masses are separated in two different 
families with the same $J$ because two different equations decouple 
for each $J$.
}
\begin{ruledtabular}
\begin{tabular}{c|ccrrrrrrrr}
n & Pse & Sca & J=1 & J=1 & J=2	& J=2 & J=3 & J=3    \\ \hline
0            &  $2 \times 10^{-1} i \over m^2$ &  $3 \times 10^{-2}  i \over m^2$ & 3.71 & 4.59 & 6.15 & 6.45 & 7.65 &  7.84   \\ 
1            &  $2 \times 10^{-3} i \over m^2$ &  $3 \times 10^{-4}  i \over m^2$ & 6.49 & 7.15 & 8.43 & 8.69 & 9.72 & 9.89   \\ 
2            &  $2 \times 10^{-5} i \over m^2$ &  $3 \times 10^{-6} i \over m^2$ & 8.76 & 9.32 & 10.45 & 10.68 & 11.61 & 11.76  \\ 
3            &  $2 \times 10^{-7} i \over m^2$ &  $3 \times 10^{-8} i \over m^2$ & 10.77 & 11.27 & 12.30 & 12.51 & 13.38 & 13.52   \\  
4            &  $2 \times 10^{-9} i \over m^2$ &  $3 \times 10^{-10} i \over m^2$ & 12.61 & 13.08 & 14.05 & 14.25 & 15.12 & 15.26   \\  
\end{tabular}
\end{ruledtabular}
\end{table}

\section{Conclusion, including temperature effects}

Assuming a confining potential, the mass $M$ spectrum of mesons is 
studied in the true chiral symmetry breaking vacuum and in the 
unstable vacuum where chiral symmetry restoration occurs. The 
only parameter is the strength $K_0$ of the potential.
Chiral models have the same number of meson states in the spectrum 
as the normal quark model. The mass splittings can de related,
as usual, to spin-tensor potentials. In the limit of vanishing
constituent quark masses, all spin-dependent potentials are
quite simple, proportional to ${K_0}^3/k^2$.

In the chiral limit the mesons suffer small $M/ K_0$ 
changes from one vacuum to the other, except 
for the $J=0$ pseudoscalars and scalars. All the
$J=0$ mesons, including all possible radial excitations,
are transformed in tachyons with infinite imaginary masses,
when the true vacuum is replaced by the chiral invariant vacuum.
An detailed analytical proof and a precise numerical
study of the tachyons are also presented here.
 
However, before moving to the conclusions, these beautiful
mathematical results should be matched with our knowledge
the deconfined phase of QCD. 

My first comment concerns the calibration problem of any chiral
symmetric model. The Sigma Model, the Nambu and Jona-Lasinio
model and Chiral Lagrangian estimations are not confining and thus
are not expected to address correctly hadrons with spin, angular or radial
excitations. The present model is adequate to study the angular or
radial excitations of hadrons, and in this sense it already upgrades
previous estimations of the meson spectra in the chiral restored vacuum.
Nevertheless the present density-density interaction 
suffers from uncalibrated spin-tensor potentials. 
But I submit that the under development chiral invariant 
quark models with a confining funnel interaction
\cite{Bicudo_KN,Llanes-Estrada_thesis} 
a vector interaction
\cite{Llanes-Estrada_hyperfine,Bicudo_scapuz},
or long range scalar interactions 
\cite{Bicudo_scalar,Villate},
can be correctly calibrated.
Nevertheless, for a qualitative study, the present density-density
harmonic confining interaction should be sufficient, since
PB and Nefediev
\cite{Bicudo_rep}
have shown that this interaction has similar mass gap solutions 
to the other possible confining potentials in Coulomb gauge QCD.

My second comment concerns the parameters of the present model.
The potential strength $K_0$, the dominant scale of the present study,
is expected to change from the ordinary QCD vacuum to the deconfined
phase of QCD. This is quite important because the meson masses scale
with $K_0$.

Mys first conclusion concerns corrections due to the current quark mass. 
The light current quark mass is small but not vanishing. The $u$ or $d$
quarks correspond to and increase the chiral limit quark constituent mass by 1\% to 2\% of $m_c$, 
while the $s$ quark amounts to increase the chiral limit quark constituent mass by
up to 50\%. For instance in the true vacuum the $s$ constituent quark mass is of the
order of 1.5 $m_c$, while in the chiral restored vacuum the $s$ constituent quark
mass is of the order of 0.5 $m_c$. These simple factors are sufficient to estimate
from the Figs. \ref{light-light PS},
\ref{light-light VA}, \ref{heavy-light PS} and \ref{heavy-light VA},
the masses of the vectors $\rho$, $\omega$ or $\phi$, or of the pseudoscalar
and vector $D$ and $D_s$, relevant for the new di-muon measurements of NA60. 
In the light-light systems, with a $u$ or $d$ quark and a $\bar u$ or $\bar d$ 
antiquark, the number and the imaginary mass of pseudoscalar and scalar tachyons 
are not infinite, nevertheless they are very large. 

Interestingly, in Fig. \ref{light-light VA} the vector meson has real mass for
zero quark masses, but for a small mass the vector meson is a tachyon.
Thus it is possible that the $\rho$ meson, or the $\omega$ meson,
simply disappear in the chiral restored vacuum. Because the quark
mass interval, where the vector meson is a tachyon, is quite small,
it is plausible that the $\rho$ meson and the $\omega$ meson may have
a different tachyonic behaviour, although the present study cannot
explore the differences between the $\rho$ and the $\omega$.
Notice that the NA60 collaboration saw differences between the production rate
of the $\rho$ and the $\omega$
\cite{NA60_1,NA60_2,NA60_3,NA60_4,NA60_5}, 
but this may also be due to $\rho$ interactions with $\pi$ at the periphery
of the deconfined QCD bubble
\cite{Dobado}.

My second conclusion is that the chiral invariant vacuum is
too unstable to be reached, unless confinement is lost.
This is clearly signalled by the infinite, (or very large) 
number of infinite (or very large) imaginary mass of 
tachyons in the pseudoscalar and scalar channels. This extreme
unstability confirms a result of Le Yaouanc et al.
\cite{Yaouanc2}, 
who studied the deconfinement transition, using the present 
confining potential, and concluded that the transition does 
not occur for any finite temperature. 
Therefore a change in the potential must happen
before the chiral restoration transition occurs. This also
confirms the lattice QCD simulations initiated by
Kogut, Wyld, Karsch and Sinclair
\cite{Kogut,Narayanan,Karsch},
and the Schwinger-Dyson calculations initiated by
Bender, Blaschke, Kalinovsky and Roberts,
\cite{Bender,Roberts}
who also found a restoration of chiral symmetry coincident with 
the loss of confinement at temperatures of the order of 150 MeV.

The third conclusion of this paper is that all the meson masses 
are much smaller in the high temperature chiral invariant vacuum, than 
they are in the low temperature symmetry breaking vacuum. 
Notice for instance that the apparently constant vector and axialvector masses 
of Fig. \ref{heavy-light VA} are proportional to the potential strength $K_0$,
thus they decrease when the potential strength decreases.
This is an educated conclusion, based on Lattice QCD simulations of the dependence
of the confining potential with temperature and also with dynamical fermions.
when confinement is lost,
\cite{Karsch}
at temperatures of the order of 150 MeV, the strength of the potential is also decreased. 
These two effects are necessary for chiral symmetry restoration. 

Assuming these two changes, both in shape and strength of the potential, 
the spectra computed in this paper can be reinterpreted.
Assuming that confinement disappears, the infinite number of infinite imaginary 
mass tachyons go away. Moreover, a smaller strength of the potential is
also necessary to remove any tachyon in the chiral symmetric vacuum.
Then the chiral symmetric vacuum is the only and true vacuum. Notice that, for
light quarks, the largely dominant scale, including the scale ruling the constituent
quark mass, is the strength of the potential. All the spectra are
proportional to the strength of the potential, see Figs  \ref{light-light PS},
\ref{light-light VA}, \ref{heavy-light PS} and \ref{heavy-light VA}. Then,
with a much weaker potential, the masses and widths of any possible mesons
are much smaller (except for the contribution of the heavy quark mass, say
the $\bar c$ mass in $D$ or $D_s$ mesons)
than the masses of ordinary mesons listed by the Particle Data Group
\cite{RPP}.  
Thus the vector mesons identified by the NA60 collaboration, with masses
close to the ordinary masses, are not expected to be probed inside the
deconfined phase of QCD, where all mesons, if any, are much lighter.
 
\acknowledgements

PB tanks Jo\~ao Seixas for extremely motivating discussions
on the results of NA60 and on lattice QCD potentials.

%bbbbbbbbbbbbbbbbbbbbbbbbbbbbbbbbbbbbbbbbbbbbbbbbbbbbbbbbbbbb
%bbbbbbbbbbbbbbbbbbbbbbbbbbbbbbbbbbbbbbbbbbbbbbbbbbbbbbbbbbbb
%bb
%bb
%bbbbbbbbbbbbbbbbbbbbbbbbbbbbbbbbbbbbbbbbbbbbbbbbbbbbbbbbbbbb
%bbbbbbbbbbbbbbbbbbbbbbbbbbbbbbbbbbbbbbbbbbbbbbbbbbbbbbbbbbbb

\end{document}